# The formation and evolution of large- and superlarge-scale structure in the universe - I: General Theory

Andrei G. Doroshkevich,[1,2] Richard Fong,[3] Stefan Gottlöber,[4] Jan P. Mücket[4] and Volker Müller[4]

[1] *Keldysh Institute of Applied Mathematics, Academy of Sciences, 125047 Moscow, Russia*
[2] *Theoretical Astrophysics Centrum, Blegsdamsvej 17, DK-2100 Copenhagen O, Denmark*
[3] *Department of Physics, University of Durham, South Road, Durham, DH1 3LE*
[4] *Astrophysical Institute of Potsdam, An der Sternwarte 16, Potsdam, D-14482 Germany*

**ABSTRACT**
We present a new quantitative theory, based on the Zel'dovich approximation, to provide an approximate description of the evolution of structure. Thus, together with that for the creation of structure, this can be seen as an essential step forward towards a complete theory for the formation and evolution of structure in the universe. We refine earlier work for the comoving characteristic scale of superlarge-scale structure (SLSS), allowing then a well-defined expression which can also be applied to presently popular power spectra with a Harrison-Zel'dovich form at large wavelengths for the initial density inhomogeneities in the universe. The scale of SLSS characterizes the average cell size of the cellular structure induced by the gravitational potential in the distribution of galaxies and of dark matter. New also is the theory for the formation and evolution of large-scale structure (LSS). We show then that for the usual assumption of Gaussian initial conditions, the creation of structure occurs rapidly and can, thus, be thought of as a 'phase transition' from a structureless universe to one with structure. For more recent epochs, the scales of LSS are expected to correspond to the mean separations of filaments or chains of galaxies at those epochs. We apply the theory to some initial power spectra and successfully show that the theoretically defined characteristic scale for LSS does also correspond to the 'physical measurement' of the 'mean separation' of dark matter LSS elements in the corresponding N-body simulations. This supports our claim for a new theory for the *quantitative* description of the formation and evolution of large-scale structure in the universe. Additionally, the theory can then quantify the practical limitations of simulations. We discuss, finally, the observational possibilities of the physical measurement of large-scale structure in the Universe.

# 1 INTRODUCTION

Large-scale structure in the Universe is one of the main sources of criteria needed for addressing the problem of the origin of the observed distribution of galaxies, and how it forms and evolves is crucial to our understanding of the many properties of galaxies and of the inter-galactic medium. Observationally, there has been over the past decade a tremendous growth in the acquisition of data on the spatial distribution of galaxies. At the same time, theoreticians have developed effective methods for the numerical simulation of the formation of large-scale structure in an expanding universe and there is now quite a rich collection of simulations with various power spectra, dark matter (DM) compositions and computational boxes. There is, thus, already a broad base of methods and observations for studying the main physical properties of the formation and evolution of structure and, in particular, for the theoretical description of the physical measures that characterise these properties.

It has been through the Zel'dovich approximation (Zel'dovich 1970) that the general properties of the formation and evolution of structure in the universe due to the physical action of gravitational instability are now so well understood; we shall briefly review these properties in Section 2. The present epoch is then seen to belong to the intermediate phase of structure evolution, following a phase in which structure is first created. In this intermediate phase the universe is permeated by a random network of 'filaments'. However, it is further modulated by a random cellular structure on very large scales induced by the gravitational potential. This naturally gives rise to 'voids' inhabited by weak filaments or chains of galaxies within the cells of this structure, with rich filaments of galaxies inhabiting the surfaces of this random cellular structure. In a preliminary investigation based on pencil beam galaxy redshift surveys, we found the mean distance between filaments to be $\sim 10\text{--}15\,h^{-1}$ Mpc and $\gtrsim 50\,h^{-1}$ Mpc for the mean size of 'voids' (Buryak, Doroshkevich & Fong 1994, hereafter BDF). Hence, we called galaxy filaments elements of large-scale structure or LSS and cells, the cell surfaces in particular, elements of superlarge-scale structure or SLSS.

In summary, the present spatial distributions of matter and of galaxies have clearly arisen from the random initial perturbations through the action of several physical factors. Of particular concern in this paper will be the effects of the non-linear transformation of the velocity field, leading to the formation of LSS, and its modulation by the gravitational potential field to produce SLSS, the latter effect being enhanced for the galaxy distribution by the 'thresholding' process for galaxy formation and by secondary ionisation (see, also, the discussion in Section 2). Other physical effects concern the dynamical evolution of the DM component within a Zel'dovich pancake and the thermal, chemical and dissipational evolution of the gaseous component.

Thus, formally, the process of structure formation and evolution can be seen as due to the non-linear transformation of the initial velocity field, which is directly calculable from the initial power spectrum (see, e.g., Demiański & Doroshkevich 1992, hereafter DD, for a more detailed discussion). An important fundamental result is that the transformation is then a *deterministic* process, being simply gravitational, and the *random* character of the structure is simply a reflection of the stochasticity of the initial conditions. In particular, this implies that the physical parameters characterizing the structure, such as the mean cell size, will be closely related to the parameters of the initial gravitational potential field and can, in principle, be formulated in terms of the power spectrum, $p(k)$, where $k$ is the comoving wave number. The theoretical challenge then is to define necessarily statistical functions that will describe well these physical parameters of large-scale structure. Over the years, expressions for various parameters have been put forward, providing some important results (see, e.g., Peebles 1993, Efstathiou et al. 1988, Ostriker & Sato 1990, Klypin & Melott 1992 and Kofman et al. 1992, hereafter KPSM).

We have, therefore, identified the main tasks for a *quantitative* theory of the formation and evolution of large-scale structure in the universe as being the following:

1. To develop a method for the quantitative description of the spatial distribution of matter.
2. To clearly identify the stochastic and deterministic elements in the description provided by the theory.
3. To extract and describe quantitatively the dominant physical factors affecting the evolution of the spatial distribution of matter.

Buryak, Demiański & Doroshkevich (1992) and BDF have provided a solution to the first two problems. Their method of analysis involves the transformation of the three-dimensional spatial distribution of 'particles' to a sequential distribution of structure elements, namely filaments and sheets, along a random straight line. The method is, thus, well suited to the analysis of pencil beam galaxy redshift surveys, for which it was originally designed, and we shall also use it here for the analysis of N-body simulations. In analogy to the similar method of sampling geological structure, we shall simply call the BDF method of analysis the *'core-sampling'* method. We showed in these papers that the *stochastic* factor appears as an uncorrelated Poisson-like spatial distribution of the structure elements along the 'core samples', while the *deterministic* effects of the physical factors manifest themselves in the evolution of the characteristic measures, i.e. the mean or average values of physical parameters of these random distributions, such as the mean separations of structure elements.

In this paper, we propose a new quantitative, albeit approximate, description of the evolution of structure and relate it to that for the formation of structure, making an essential step forward towards a complete theory. Again our starting point is the Zel'dovich approximation, the great potential of which has been repeatedly demonstrated. We show that the evolution of LSS can be described by a few approximate expressions involving simply moments of the power spectrum. In particular, we shall establish the relationship of characteristic measures of LSS and SLSS and of their evolutionary behaviour with the power spectrum. Some aspects of the formation of SLSS has already been discussed by DD and Buryak, Demiański & Doroshkevich (1993, hereafter BDD).

Section 2 provides a brief review of the qualitative picture of the formation and evolution of structure in the universe. In Section 3 we present our approximate theoretical methods for estimating the characteristic scales of LSS and SLSS, finding also the time dependence of the LSS scale. We obtain then the results for some power spectra of present interest for the initial density inhomogeneities in the universe. We show in Section 4.1 that these theoretical definitions of the characteristic scale of LSS do indeed correspond to 'measurements', using the 'core sampling' technique, of the 'mean separation' of LSS elements in the dark matter distribution in N-body simulations. Section 4.2 then makes use of the theory to delineate some of the limitations of N-body simulations now and in the near future due to their necessarily limited dynamic range. However, we leave to Paper II (Doroshkevich et al. 1995) the detailed discussion of the analysis used to estimate these characteristic scales in N-body simulations. We finally summarise our conclusions and discuss our results in the context of observations in Section 5

## 2 RESULTS OF THE QUALITATIVE THEORY

The most natural theory for the formation of structure in the Universe is that of the growth of small inhomogeneities in the early universe through gravitational instability. With present ideas regarding the nature of matter in the Universe, the theory is clearly applicable to the dark matter content of the Universe. We shall also make the generally accepted hypothesis that galaxies will preferentially be formed in regions of 'significant' mass over-densities. The theory is certainly consistent with observations, with perhaps the strongest evidence being provided by the COBE results for the fluctuations in the cosmic microwave background (see, e.g., Dekel (1994) for a recent review discussion).

The problem of the formation of structure has in the main been solved by Zel'dovich's approximate theory (1970), by its development for the 'statistical' description of the formation of structure (Doroshkevich 1984, 1988, DD) and through the ideas of 'adhesion' theory (KPSM). Generally, structure is formed through rapid collapse along one axis to form a Zel'dovich 'pancake', with less rapid collapse along a second orthogonal axis, to form a 'filament', that is, creating a 'ridge' of higher over-density within the pancake. It is most clearly studied through the velocity field, as this obviously describes how the fluid particles in the universe stream towards each other to form the structure elements of pancakes and filaments. In these terms, the structure elements are formed by the coalescence of particles between minima of the velocity potential, involving deeper neighbouring minima with time. In particular, structure elements merge and only the deeper minima survive. Thus, the general picture that emerges is one of the formation of pancakes and filaments of matter. They then evolve not only by the continued accretion of matter, but also by merging together and by a general motion of matter along the filaments towards the branch points of this 'general network structure'. The branch points are another important class of large-scale structure; they are the sites for the formation of the richest clusters of matter. We refer the reader to Shandarin & Zel'dovich (1989) for an excellent review of much of this work. However, although these processes have now been simulated, their theoretical description is still at a rather primitive stage.

Additionally, this random network structure is 'modulated' by the gravitational potential field. It can easily be seen from Poisson's equation that relative to the initial density inhomogeneities the gravitational field is dominated by power on large wavelengths. Thus, growth will proceed more rapidly in the contracting over-dense regions of negative potential, with the least rapid growth in the expanding under-dense regions of positive potential. This will create a cellular structure with rich DM pancakes lying in the surfaces of cells and weak DM pancakes within cells. The mean cell size is expected to be $\gtrsim 50\ h^{-1}$ Mpc, as equations (3.3a–d) below show. With such a large scale, we term this as the scale for superlarge-scale structure, SLSS. In contrast, the scale for the mean separation of Zel'dovich DM pancakes is $\sim 10\ h^{-1}$ Mpc (Section 3.5), and this we then call the scale for large-scale structure or LSS. In particular, there are thus two types of structure, one of which, namely LSS, is the result of the random physical motion of matter, whereas the other, i.e. SLSS, relates to the large scale modulation of the velocity (and density) field by the spatial distribution of the perturbations of the gravitational potential.

As for the distribution of galaxies, our natural assumption would imply that galaxies mainly trace the high density ridges of dark matter pancakes to produce chains or filaments of galaxies. This can be seen directly from 'core-sampling' analyses of N-body simulations of 'galaxy' distributions (Doroshkevich, Fong & Makarova 1995, hereafter DFM). However, it should be remembered that, although 'plausible' assumptions or ansatz are made in dynamical simulations, it is still unknown as to how galaxies actually form in these filaments. Thus, such N-body models should at present be seen as providing qualitative insights rather than quantitative predictions for the formation and evolution of structure in the universe.

In particular, the simulations have not yet found a way of implementing the physical process of secondary ionization. The results of observations made for the Gunn-Peterson effect (Gunn & Peterson 1965; Schneider, Schmidt & Gunn 1991; Jakobsen et al. 1994) show that the inter-galactic medium is strongly ionized. It is expected that this occurs through the UV radiation from the first objects to be formed, that is, the QSOs, AGNs, young stars, etc., in the first structures to form. This will then tend to prohibit the formation of galaxies that may 'normally' have taken place later in the weaker filaments. In particular, this process will turn under-dense regions of dark matter into the 'voids' that we see in the galaxy distribution. Because of the stochastic nature of the initial conditions, such 'voids' are not expected to be completely empty of galaxies; the few galaxies seen in 'voids' are expected to be associated with weak dark matter pancakes. 'Voids' are, thus, under-dense regions in the distribution of galaxies, but much more highly under-dense than for the dark matter there. Furthermore, as galaxies clearly do not trace continuously large-scale structure, we are then naturally led to consider that the distribution of galaxies forms a 'broken cellular structure' (BDF). A fuller discussion of the problem of dark matter in 'voids' can be found in Fong, Doroshkevich & Turchaninov (1995).

An important consequence of such physical processes is that the biasing of galaxies relative to dark matter will then depend on environment and will, thus, be a non-trivial function of position. This emphasizes the need to treat the results of N-body simulations with some caution, especially in their use for the interpretation of the observational results for galaxies, QSOs, etc.

The qualitative picture provided by gravitational instability theory is strengthened, indeed even suggested, by the increasing number of galaxy redshift surveys, some of which are now reaching $\sim 500\, h^{-1}$ Mpc in depth, confirming, for example, the general existence of large-scale structures comparable to the 'Great Wall' and the 'Boötes Void'. It is also particularly interesting to note the strong filamentary structure in the 'Great Wall', which can clearly be seen in Fig. 5(a) of Ramella, Geller & Huchra (1992), since the theoretical expectation is that the 'Great Wall' should simply be composed of cell surfaces permeated by rich filaments and rich clusters of galaxies. It thus seems that the present epoch corresponds to the second intermediate phase of structure evolution, the phase following the creation of structure and dominated by the collapse of the smaller 'voids' and the merging of LSS structure elements.

# 3. PARAMETERS AND EVOLUTION OF LSS AND SLSS

In this Section, we present a new *quantitative* theory for a description, albeit an approximate one, of the formation and evolution of LSS and SLSS for the usual case of Gaussian initial conditions, concentrating, in particular, on broad-band power spectra. As usual in linear theory, we consider the perturbations on different wavelengths to be decoupled and, thus, to evolve independently in the early universe, so that the *primordial* and *post-recombination* power spectra, $p_0(k)$ and $p(k)$, respectively, are simply linearly related,
$$p(k) = T^2(k)p_0(k).$$
Following convention, power spectra are normalized to the COBE quadrupole result (Smoot et al. 1992), so that the equation can be written without the need to include explicitly the trivial evolution of the power spectra in the linear regime. $T(k)$ is naturally called the transfer function for the model and depends on the physical processes which affect the evolution of the early universe up to the epoch just after recombination. As we shall often refer to it in this paper, we shall henceforth denote the post-recombination power spectrum as simply the power spectrum.

## 3.1 The parameters of SLSS.

We first briefly recall the main ideas of DD and BDD for describing the typical scale of SLSS as an 'observable' manifestation of the influence of the gravitational potential perturbations on the process of LSS formation.

In terms of the power spectrum, $p(k)$, a typical scale, $L_0$, can be defined as

$$L_0^2 \equiv 3\pi^2 \frac{\int_0^\infty k^{-4}p(k)k^2 dk}{\int_0^\infty k^{-2}p(k)k^2 dk}, \qquad (3.1)$$

where the coefficient of $\pi^2$ in (3.1) arises from the assumption of Gaussian initial perturbations. $L_0$ is then the typical distance between points of intersection of a random straight line with the equipotential surface $\varphi = \bar{\varphi}$, where $\varphi$ and $\bar{\varphi}$ are the actual and mean values of the gravitational potential, or, equivalently, the velocity potential, $\Phi$, since in linear theory $\varphi = \frac{3}{2}H\Phi$. As usual, we set the potential such that $\bar{\varphi} = 0$. However, although equation (3.1) holds for many power spectra, $L_0$ is logarithmic divergent for popular spectra that have resulted from a Harrison-Zel'dovich primordial spectrum. Thus, we require a more refined definition of this scale than was given by DD and BDD.

Physically, the problem is caused by the presence of very long wavelengths, up to infinity in principle, allowing the presence then of intersections $\to \infty$. Thus, the mean value of the intersections of a line with $\varphi = 0$ is undefined for such spectra. However, the zero of potential has, of course, no physical content and we should more properly be considering the potential difference between any two points rather than its absolute value at any one point. It is this crucial aspect of potentials that provides us with a more meaningful definition of $L_0$.

As it is the mean free path between SLSS that is of interest here (see the discussion in Section 1 about the 'core-sampling' method) and recalling also the spatial homogeneity of the universe, this problem is clearly best solved by considering the statistics of the gravitational potential along a random straight line. Given a random straight line, we are at liberty to choose any point on it as the origin. Take then a point on the line at the distance coordinate, $r$, and consider the potential difference, $\psi_\tau$, with a second point on the line and a further distance $\tau$ away, i.e. at the coordinate $r + \tau$,

$$\psi_\tau(r) = \varphi(r + \tau) - \varphi(r).$$

For fixed $\tau$, this then defines a random function of a single parameter, the coordinate $r$, of the 'reference point' from which the potential difference is to be evaluated.

The theory of random processes (e.g. Rice 1954) can now be used to find the number density of these 'reference points' for which $\psi_\tau = 0$, i.e.

$$n(\tau) = \lim_{\mathcal{L} \to \infty} \frac{N(\tau)\big|_{\psi_\tau = 0}}{\mathcal{L}},$$

where $N(\tau)|_{\psi_\tau=0}$ is the number of points on a random straight line of length, $\mathcal{L}$, for which $\psi_\tau = 0$. This then *allows* us to define a mean distance, $L_\tau$, between 'reference points' for which $\psi_\tau$ vanishes as

$$L_\tau \equiv \frac{1}{n(\tau)}.$$

For SLSS, we expect to find that $L_\tau$ would, in fact, correspond to $\tau$ itself, as then the sections, $(r, r + \tau)$, of the line, for which $\psi_\tau(r) = 0$, are either separated or overlap by only small amounts in distance. Thus, we can solve our physical problem analytically using the theory of random processes if we simply determine the 'label', $\tau$, for which the random function of $r$, $\psi_\tau(r)$, leads to a mean distance, $L_\tau$, equal to the 'label' itself, i.e. for which

$$L_\tau = \tau.$$

The characteristic scale of SLSS, $L_0$, can then be *defined* as this mean distance. Mathematically, this just means solving the implicit equation

$$L_0 = \lim_{\mathcal{L} \to \infty} \frac{\mathcal{L}}{N(L_0)\big|_{\psi_{L_0}=0}}.$$

Similar problems in a cosmological context have previously been studied by Doroshkevich (1970), Bardeen et al (1986, hereafter BBKS) and Doroshkevich & Kotok (1990).

Thus, instead of the explicit equation (3.1), we now have the following implicit equation for $L_0$:

$$L_0^2 = \pi^2 \frac{s_{11}}{s_{22}}(1 - \kappa^2), \tag{3.2}$$

where

$$\kappa^2 = \frac{s_{12}^2}{s_{11} s_{22}},$$

with

$$s_{11} = \frac{1}{\pi^2} \int_0^\infty p(k)\left(1 - \frac{\sin kL_0}{kL_0}\right) k^{-2} dk,$$

$$s_{12} = \frac{1}{\pi^2} \int_0^\infty p(k) \sqrt{\frac{\pi}{2kL_0}} J_{3/2}(kL_0) k^{-1} dk,$$

$$s_{22} = \frac{1}{3\pi^2} \int_0^\infty p(k) \left(1 + \frac{\sin kL_0}{kL_0} - \sqrt{\frac{2\pi}{(kL_0)}} J_{5/2}(kL_0)\right) dk$$

and $J_j$ the Bessel function of order $j$. If now the power spectrum is such that the contributions from the region $kL_0 < 1$ to the integrals for $s_{11}$, $s_{12}$ and $s_{22}$ are negligible, then the definitions, 3.1 and 3.2, become equivalent. As we shall see in Section 4, we have just this situation with N-body simulations, where we always have to deal with a finite box size.

For the BBKS and Bond & Efstathiou (1984) CDM power spectra, which we differentiate here by $CDM^a$ and $CDM^b$, respectively, the Warm Dark Matter (WDM, BBKS), the Broken Scale Invariance ( BSI, Gottlöber et al., 1994) and CHDM power spectrum (Klypin et al, 1993, hereafter KHPR), which all follow a Harrison-Zel'dovich spectrum at large enough wavelengths, we obtain, respectively,

$$L_0^{CDM^a} = 31\, h^{-2}\,\mathrm{Mpc}, \qquad (3.3a)$$
$$L_0^{CDM^b} = 27\, h^{-2}\,\mathrm{Mpc}, \qquad (3.3b)$$
$$L_0^{WDM} = 26\, h^{-2}\,\mathrm{Mpc}, \qquad (3.3c)$$
$$L_0^{BSI} = 118\, h^{-1}\,\mathrm{Mpc} \qquad (3.3d)$$

and

$$L_0^{CHDM} = 128\, h^{-1}\,\mathrm{Mpc}, \qquad (3.3e)$$

where the $h$ dependence depends on the transfer function of the model. Clearly, the large size of these scales imply that they are essentially comoving scales, and, as in BDF, we identify these as scales for SLSS.

It should, however, be noted that these new estimates for the characteristic scale, $L_0$, of SLSS are about two times smaller than our previous estimates quoted in BDF, which were based on equation (3.1) with a longwave cut-off at the present horizon size. Consequently, this more refined definition for $L_0$ has a significant effect for the theoretical scale of SLSS resulting from power spectra derived from a Harrison-Zel'dovich primordial power spectrum.

The 'core-sampling' method of BDF allows one to determine observationally the distance, $D_{SLSS}$, between 'superclusters' and, thus, between deep troughs in the gravitational potential. We expect then that

$$D_{SLSS} \approx 2 \times L_0 \approx (50\text{--}150)\, h^{-1}\,\mathrm{Mpc}.$$

The important physical significance of such a result is discussed in BDF. However, the data available to BDF were rather limited and the observational estimation of $D_{SLSS}$ needs to be further investigated using the new galaxy redshift surveys that are now being acquired.

## 3.2 The phase of structure formation

DD and BDD also provided a description of the first phase of the evolution of structure, in which the dominant process is the creation of structure with the birth of new structure elements. In the Zel'dovich approximation, the main parameters for the evolution of structure are the amplitude of perturbations, $\sigma_\rho$, and the (comoving) characteristic scale, $r_c$. In terms of the power spectrum, $p(k)$, they are defined as

$$\sigma_\rho^2 \equiv \frac{1}{2\pi^2} \int_0^\infty p(k) k^2 dk \qquad (3.4a)$$

and

$$r_c^2 \equiv \frac{3 \int_0^\infty p(k) k^2 dk}{\int_0^\infty k^4 p(k) dk} \qquad (3.4b)$$

Clearly, these values are not defined for some of the presently favoured power spectra, for which there is not a short-wave limit. However, as we shall discuss at the end of this section, there is expected to be a natural limit to the short wave part of the spectrum corresponding to the mass of DM particles and the parameters $\sigma_\rho^2$ and $r_c$ are then directly related to this mass.

As to galaxy filamentary structure, we identify this, as did BDF, with LSS and it is the parameter, $r_c$, which then determines the spatial scale of LSS during this first phase of structure formation. However, the 'core-sampling' analysis of N-body simulations shows that for the dark matter, instead of filaments, the dominant structure elements are sheet-like with the expected characteristic scale for LSS (DFM). Thus, in such simulations, the 'galaxy' filaments are just the central high density ridges of DM 'Zel'dovich pancakes' (Section 2). To avoid confusion, we shall generally refer to these as elements of LSS.

During this epoch, the creation of LSS elements is dominant, and we can neglect any merging. It can then be shown that, for a matter dominated universe with $\Omega = 1$ and Gaussian initial conditions, the mean distance, $l_{cr}$, between elements decreases during this epoch as

$$l_{cr}^{-1} \simeq r_c^{-1} \left[ \frac{8}{3\pi^2} \sqrt{\frac{15}{7}} \left(1 + y^2/2\right) \exp\left(-y^2/2\right) \right] \leq 0.4 \; r_c^{-1}, \qquad (3.5)$$

$$y = (1+z)/(1+z_{cr}) = (1+z)\sqrt{5}/\sigma_\rho,$$

where $z$ is the redshift and $z_{cr}$ is the 'characteristic epoch' for the creation of LSS elements (Doroshkevich 1984, 1989).

The expressions for $\sigma_\rho$ and $\int_0^\infty k^4 p(k) dk$ are divergent for *scale free* CDM spectra and, thus, for such idealized spectra $r_c$ vanishes and the epoch of structure creation is at '$z = \infty$'. But, in actuality there is a cut-off at large $k$. According to current WDM models (BBKS), this cut-off is directly related to the mass, $M_{part}$, of the DM particles dominating the universe, i.e. $k_{max} \propto M_{part}(M_{part}/\Omega h^2)^{1/3}$. As is well known, for an HDM model with massive 30 eV neutrinos the epoch of structure formation is much too late; for this model $r_c = 9 \, h^{-1}$ Mpc and $1 + z_{cr} = 0.4$.

Unfortunately, the mass of the DM particles expected to dominate the evolution of the universe is at present unknown. Interestingly, this suggests that if we could observe the epoch of structure formation, $z_{cr}$, this would 'predict' the mass of DM particles within the context of present WDM theories. For example, a DM particle mass of $\sim 3$ keV could correspond to $r_c \sim 0.07\,h^{-1}$ Mpc and $z_{cr} \sim 10$ .

However, we should note that there will be some 'local' influence of the gravitational potential on the formation of structure; it will be earlier and faster in the regions of negative potential, and later and slower in regions of positive potential. As $(\delta\rho/\rho)^2$ is given by $p(k)$, Poisson's equation implies that $(\delta\varphi/\varphi)^2$ will be given by $p(k)/k^4$. The gravitational potential will then be dominated by long wavelengths, resulting in a very smooth function spatially. Thus, this influence of the potential will lead to a very large-scale modulation of $z_{cr}$ (DD). Furthermore, secondary ionization of the intergalactic medium due to UV radiation from the young stars and QSOs in the first structures to be formed will inhibit later weaker structures from forming galaxies, enhancing the effect of this modulation on the distribution of galaxies compared to its effect on the DM distribution (Section 2). It is, of course, this effect that gives rise to SLSS, that is, to superlarge-scale structure, and thus to the possible presence of such a scale as those of equations (3.3) in the galaxy distribution.

### 3.3 The accretion of matter onto LSS elements.

The results of Section 3.2 imply extremely early structure formation with an extremely small characteristic scale, $r_c$, for broad band power spectra, such as those of the CDM models. However, further evolution of the structure is mainly dictated by the initial velocity field, which has neither divergencies or any other peculiarity – discontinuities occur only for the derivative of velocity. We can, therefore, use quite standard methods for the theoretical investigation of the evolution of structure.

As is well-known, the evolution of structure is determined by two main processes (see, e.g., the studies of adhesion theory and the results of numerous N-body simulations). At the period of structure formation and for a short time afterwards, it is just the accretion of matter onto the structure elements that is most important. Indeed, only with accretion are the 'caustics' of the theory transformed into real structure elements. So, first of all, we need to provide some approximate quantitative description of this accretion of matter. The second very important process is the motion, intersection and merging of structure elements, which we shall be treating in Section 3.4. The accretion process was considered by Doroshkevich & Shandarin (1978) and we will follow their treatment for our description of the process for broad band power spectra. It is also based on Zel'dovich's approximate theory of gravitational instability (Zel'dovich 1970).

In the Zel'dovich approximation (see, e.g., Shandarin & Zel'dovich 1989), the Eulerian coordinate, **r**, (the proper coordinate of the actual position) is related to the Lagrangian coordinate, **q**, (the initial comoving coordinate) of the particle as

$$\mathbf{r}(\mathbf{q},t) = a(t)\,(\mathbf{q} - b(t)\mathbf{s}(\mathbf{q})) \qquad (3.6)$$

where **s** is an 'initial displacement' or 'spatial perturbation' and where, for $\Omega = 1$, $a(t) = (1+z)^{-1}$ is the expansion factor and the growth of perturbations factor, $b(t)$, is also equal to $(1+z)^{-1}$. It then follows that the 'peculiar velocity'

$$\mathbf{v} = (1+z)^{-2} H(z) \mathbf{s},$$

where $H = \dot{a}/a$ is, of course, the Hubble constant at the epoch, $z$.

The standard correlation function of the displacements

$$< \mathbf{s}(\mathbf{q}_1)\mathbf{s}(\mathbf{q}_2) > = g(q_{12}) = g(|\mathbf{q}_1 - \mathbf{q}_2|) = \frac{1}{2\pi^2} \int_0^\infty p(k) \frac{\sin kq}{kq} dk \quad (3.7)$$

is a smooth function and for a Harrison-Zel'dovich primordial spectrum it can be fitted over a wide range in $q$ by the simple relationship

$$g(q)/g(0) \approx \sqrt{1+x^2} - x, \quad x = q/l_0 > 0.1, \quad (3.8a)$$

with

$$l_0^{-2} = \int_0^\infty kT^2(k)dk, \quad (3.8b)$$

where $T(k)$ is the transfer function of the CDM model. Thus, this relationship leads to a new characteristic scale, namely $l_0$, which is of the same size as the correlation radius of displacements ($r_s = 0.75 l_0$, $g(r_s)/g(0) = 0.5$), as well as being much greater than $r_c$. At first sight, this would seem to imply the accretion of matter over very large scales. However, a closer examination of this problem shows that the situation is not as simple as it seems.

Consider then a particular pancake and choose a coordinate system with origin at the centre of the pancake. Clearly, it is the motion of the matter along the direction, $\mathbf{n}$, normal to the caustic surface that results in the accretion of matter onto the pancake. The condition that the Eulerian thickness of the pancake is so small as to be negligible is then

$$\mathbf{n} \cdot \mathbf{r} = 0.$$

Thus, accretion is described by the equation,

$$\mathbf{n} \cdot \mathbf{q} = (1+z)^{-1} \mathbf{n} \cdot \mathbf{s}^*(\mathbf{q}). \quad (3.9a)$$

Here, $\mathbf{s}^*(\mathbf{q})$ is a 'conditional' displacement, i.e. it is a random variable dependent on the value of the largest eigenvalue of the deformation tensor (at $\mathbf{q} = 0$). The solution, $q_{pan}(z)$, of equation (3.9a) gives the Lagrangian thickness of the pancake, i.e. $d_{pan} = 2q_{pan}$.

Now, equation (3.9a) can be rewritten as

$$q = \mathbf{n} \cdot \mathbf{q} = (1+z)^{-1} \mathbf{n} \cdot \mathbf{s}^* = \frac{1+z_f}{1+z} \cdot \frac{q}{\sigma_0^2} \int_0^\infty F_{ac}(kq) p(k) k^2 dk, \quad (3.9b)$$

where $z_f$ is the redshift of the caustic at the position under consideration and the function $F_{ac}(x)$ is

$$F_{ac}(x) = \frac{7.5}{x^2} \left( (4 - 9x^{-2}) \frac{\sin x}{x} + (9x^{-2} - 1)\cos x \right).$$

Thus, the thickness of the pancake,

$$d_{pan} \approx 2.2 r_c \sqrt{1 - \frac{1+z}{1+z_f}}, \quad z < z_f \qquad (3.10)$$

which is close to the pancake separation, $l_{cr}$, as can be seen from equation (3.5). Thus, the accretion of matter leads also to the onset of the merging of structure elements and, thus, to an increase in the characteristic scale of LSS. Also, equation (3.10) shows that, following the formation of structure, there is a quite rapid exhaustion of the 'reserves' of matter that can condense onto structure, leading to a slowing down of the creation of further structure elements. Thus, in the following phase of structure evolution, which we discuss next, accretion and merging become the two dominant processes.

### 3.4 The second phase of structure evolution; the dissipation of structure.

In Sections 3.1 and 3.2, we followed DD and BDD to provide some very simple analytical methods for estimating the most fundamental parameters of LSS and SLSS for the phase in which structure first forms. Here, we extend these same considerations to the next phase. In this phase, the rate at which new elements are created has greatly diminished and the merging of existing elements, i.e. the dissipation of structure, has now become the dominant process, causing an increase in the distance between elements and, thus, of the characteristic scale for LSS. In the third and last phase, this random structure begins to disappear, as matter flows along the elements toward the branch points of the structure to give finally a 'point' distribution of rich clusters of galaxies, with some merging of the branch points. As noted by BDF, observations clearly show that the present epoch corresponds to the second, intermediate, phase. Thus, from the point of view of the observational possibilities, it is, probably, the more interesting phase to understand.

Structure formation 'erases' the power spectrum at short wavelengths in the sense that, once structure on these scales has formed, this part of the power spectrum plays little part in the future evolution of large-scale structure. Of course, for the resulting power spectrum strong short-wave amplitudes are generated by the non-linear processes of the inner evolution of the structure itself. But, at this stage these small scale motions are not describing the flow of particles to form structure elements, they describe instead the crossing and relaxation of particles within elements and perhaps also the merging of structure elements, accompanied by the formation of multi-stream regions, large gradients in the density and velocity of DM, and by shock waves in the gaseous component (for a more detailed general discussion, see Shandarin & Zeldovich 1989). The perturbations so generated are strongly correlated and it is essential for theoretical work to understand that the standard random phase approximation cannot then be used to describe such perturbations. Later, after relaxation, these motions have mainly been transformed into the internal motion of matter in quasi-stationary structure elements and, observationally, cause in redshift space a distortion of the regular structure (similar to the well-known 'fingers of God' effect seen in rich clusters of galaxies).

This extended stage of structure dissipation is governed by the displacement of LSS elements. Now, as the inner motions essentially cancel when averaged out, they cannot have a significant effect on the motion of the element as a whole. Thus, the evolution of LSS is determined by the power spectrum of initial perturbations at longer wavelengths. We consider then that the rate of dissipative processes at this stage is probably best characterized by the bulk velocity for structures of size comparable to the dissipative scale, $l_{dis}$, i.e. the scale at which structure elements are merging.

In summary then, the influence of the perturbations on small scales during this phase becomes insignificant for the rate at which large-scale structure evolves, as they really only describe the motion of particles *within* filaments and pancakes. Thus, the main influence is due to longer wavelength perturbations, which do describe how filaments and pancakes move as a whole. For these larger scales, the distortion of the power spectrum by non-linear processes is small and, thus, the greatest effect is simply that of the gravitational evolution of the power spectrum in the linear regime. An effective, and practical, way of studying the evolution of the characteristic scale of LSS, $l_{dis}$, is to use the usual smoothing procedure of analyses of large-scale structure (see, e.g., Coles et al, 1993), together with the Zel'dovich approximation.

To describe the evolution of structure, we need to consider the motion of the structure element *as a whole* and, so, we need to extend the Zel'dovich theory to include the comoving displacement and velocity of structure elements and of systems of structure elements. Obviously, the present discussions in the literature of observed 'bulk' motions are closely related.

We could, as is commonly done, seek to define a 'bulk' displacement, $\mathbf{s_{bulk}}$, or, equivalenly, bulk velocity,

$$\mathbf{v}_{bulk} = (1+z)^{-2} H(z) \mathbf{s}_{bulk},$$

of the particles within a 'bulk volume' of size $R_S$. This is most conveniently done using the Gausssian filter, whose Fourier transform is

$$S(k, R_S) = \exp(-R_S^2 k^2), \tag{3.11}$$

as the smoothing function. Another 'standard' choice of smoothing is to use the 'top hat' filter (see, e.g., Shandarin, 1993). Clearly, the Gaussian filter avoids the hard edges of the 'top hat' filter, which also produces oscillations about zero in the 'wings' of its Fourier transform. Using the Gaussian filter then,

$$\mathbf{s}_{bulk}(\mathbf{q}, R_S) = \frac{R_S^{-3}}{(2\pi)^{\frac{3}{2}}} \int \mathbf{s}(\mathbf{q}') e^{-\frac{(\mathbf{q}-\mathbf{q}')^2}{2R_S^2}} d^3 q'.$$

and the mean comoving displacement of matter elements of size $R_S$, $\sqrt{<s_{bulk}^2(R_S)>}$, is given by

$$<s_{bulk}^2(R_S)> = \frac{1}{2\pi^2} \int_0^\infty p(k) S(k, R_S) dk. \tag{3.12}$$

Our results for the function, $\bar{s}_{bulk}(R_S) \equiv \sqrt{<s_{bulk}^2(R_S)>}$, obtained by simply numerically evaluating the integral for the CDM power spectrum of BBKS, is compared in Fig. 1 with direct estimates of the bulk velocity in dynamical numerical simulations based on this same spectrum (Eke et al. 1995). As might be expected, we find good agreement of our analytical approach with the estimates from simulations.

However, although the smoothing has been made in Lagrangian space, this 'bulk' displacement, $\bar{s}_{bulk}$, is not really characteristic of the essential motion leading to the merging of pancakes. For the dissipation scale it is clearly the component of the bulk motion perpendicular to the pancake that is 'dissipated' during pancake merging. The matter motion along the other directions is conserved over the merging process, after which it is gradually transformed as it evolves into the inner 'fluid' motion of the merged structure element; it describes then the deformation and disruption of structure elements. This process is completely identical to the well known transformation of the motion of the gaseous component into the shock wave surrounding a Zel'dovich pancake, when the normal component of the velocity is dissipated into heating up the gas, but with the tangential component retained to form the secondary vortex velocity field within the pancake (Doroshkevich 1973, Zel'dovich & Novikov 1983). Therefore, we shall consider that the dissipation scale is actually given by the criterion that the relative (one-dimensional) motion of two structure elements is such as to make them meet each other half way, with, consequently, the one-dimensional dissipation of their bulk motion. In particular, it is *one-dimensional* smoothing that is clearly required.

As the one-dimensional velocity, $v_1$, is just $\partial \Phi / \partial q_1$, then the mean one-dimensional displacement velocity of matter elements of cross-sectional size, $2R_1$, is just

$$< v_{1,bulk}^2 > = H^2 (1+z)^{-4} \frac{1}{(2\pi)^3} \int \frac{p(k)}{k^4} k_1^2 \exp(-k_1^2 R_1^2) d^3k,$$

where we have used one-dimensional Gaussian smoothing for the dissipated component, $k_1$. This then allows us to calculate from the power spectrum the corresponding mean comoving distance, $\sqrt{< d_{1,bulk}^2 >}$, travelled by such structure elements, since, from equation (3.6),

$$< d_{1,bulk}^2 > = \frac{< s_{1,bulk}^2 >}{(1+z)^2},$$

with

$$< v_{1,bulk}^2 > = H^2 (1+z)^{-4} < s_{1,bulk}^2 > . \tag{3.13}$$

We now consider that pancakes merge when both their lateral size and their mean *relative* distance travelled become approximately equal, i.e. become by definition the dissipative scale, $l_{dis}$. Hence, we have chosen, as our criterion for the merging of two neighbouring elements with *uncorrelated* displacements, $\sqrt{< s_{1,bulk}^2 >}$,

$$R_1 = l_{dis} \quad \text{and} \quad l_{dis}^2 = 2 < d_{1,bulk}^2 > = \frac{2}{(1+z)^2} < s_{1,bulk}^2 > .$$

Of course, our arguments here have simply been heuristic ones and the final justification for these relationships for $l_{dis}$ can really only come with the successful testing of our theory using N-body simulations (Section 4). Most importantly, the criterion now provides us with a possible *ansatz* for the dependence of the dissipative scale not only on the power spectrum, but also on redshift, i.e. its evolutionary behaviour. Using it we obtain the following implicit equation for $l_{dis}$:

$$l_{dis}^2 = \frac{2}{(1+z)^2} \frac{1}{(2\pi)^3} \int_{-\infty}^{\infty} \frac{k_1^2}{k^4} p(k) \exp(-k_1^2 l_{dis}^2) d^3k. \tag{3.14}$$

Thus,
$$l_{dis}^2 = \frac{2}{(1+z)^2}\frac{1}{2\pi^2}\int_0^\infty p(\kappa)G(\kappa l_{dis})d\kappa,$$
where
$$G(\zeta) = \zeta^{-3}\int_0^\zeta u^2\exp(-u^2)du\ .$$

Although our approach here is qualitatively similar to that of the adhesion approximation, the essential difference is that in the adhesion approximation quantitative results can only be obtained through N-body simulations, whereas here we have been able to obtain for the first time a *quantitative theory* for the description of the evolution of structure in the universe.

It is also possible to attempt to investigate this problem by considering the deformation tensor and other fluid dynamical elements of the Zel'dovich theory. However, in our approach we are concerned more with the *discrete* distribution of structure elements and their velocities, which are determined both by their mutual gravitational interaction and by the process of the intersection and merging of structure elements. Thus, we have here to use a statistical mechanical description of pancake motion, rather than a fluid dynamical one. Of course, for large enough smoothing scales, $R_s \gg l_{dis}$, the two approaches become equivalent.

Finally, we define the characteristic scale of LSS for *both* the first and intermediate evolutionary phases as simply
$$l_{LSS} = max(l_{cr}, l_{dis}) \qquad (3.15)$$

### 3.5 The characteristic scale of LSS

Let us consider broad-band power spectra with a Harrison-Zel'dovich primordial spectrum,
$$p_0 = Ak, \quad \text{with} \quad A = 1.2\pi^2 Q_2^2 R_h^4,$$
where $R_h = 6000\,h^{-1}$ Mpc is the present horizon size and $Q_2 = (17/2.73)\times 10^{-6}$ is the amplitude of the quadrupole fluctuations of the cosmic microwave background radiation as measured by COBE (Smoot et al. 1992). Equation (3.14) then gives for the asymptotic dependences of $l_{dis}$ on $z$,

$$l_{dis}\to 0 \;\Rightarrow\; l_{dis}^2 \approx \frac{1}{3\pi^2(1+z)^2}\int_0^\infty p(k)dk = \frac{Al_0^{-2}}{3\pi^2(1+z)^2} = 0.375\,l_{max}^4 l_0^{-2}(1+z)^{-2}$$

and

$$l_{dis}\to\infty \;\Rightarrow\; l_{dis}^2 \approx \frac{A}{\pi^2(1+z)^2}\int_0^\infty k e^{-k^2 l_{dis}^2}dk = \frac{Al_{dis}^{-2}}{2\pi^2(1+z)^2} = \frac{9}{16}l_{max}^4 l_{dis}^{-2}(1+z)^{-2},$$

where the value $l_0$ is defined by equation (3.8b) and

$$l_{max}^2 = \sqrt{\frac{16}{15}}Q_2 R_h^2, \quad \text{giving} \quad l_{max}\approx 15.2\,h^{-1}\,\text{Mpc}.$$

We, thus, propose to use the very simple form

$$l_{dis}^2 \approx \frac{(1+z_t)^2}{(1+z)^2} \frac{S_0^2}{2 + l_{dis}^2/S_0^2}$$

as a smooth interpolation between these two regions. Solving this quadratic equation in $l_{dis}^2$, we find that it has only one real solution for $l_{dis}$,

$$l_{dis}^2 \approx S_0^2 \left[\sqrt{1+\zeta^2} - 1\right] = \frac{3\, l_{max}^4 (1+z)^{-2}}{4\left[\sqrt{l_0^4 + l_{max}^4 (1+z)^{-2}} + l_0^2\right]}, \quad (3.16)$$

with

$$S_0^2 = 0.75\, l_0^2, \quad \zeta = (1+z_t)/(1+z) \quad \text{and} \quad 1+z_t = \sqrt{\frac{16}{15} Q_2 R_h^2\, l_0^{-2}} = l_{max}^2/l_0^2\,.$$

For the CDM model the difference between using (3.14) and (3.16) amounts to no more than 15%. We propose then to use (3.16) in our general theoretical discussions, as well as using it as a simple way of characterising the different models of structure evolution.

Thus, for decreasing redshifts, the dissipation scales for all the models with broad-band power spectra are then all rather similar, with values close to $0.87\, l_{max}/\sqrt{1+z}$. It is important to note that the value $l_{max}$ for the Harrison-Zel'dovich primordial power spectrum depends on the amplitude of perturbations (and, thus, on the value of the quadrupole component $Q_2$ in equation (3.16)) and on the present horizon, $R_h$, only. The specific shape of the transfer function, which affects the value $l_0$, plays a more important role during the earlier period of structure evolution, $z \gg z_t$, when $l_0 \gg l_{max}/\sqrt{1+z}$. It can be seen from equation (3.16) that $(1+z_t)$ defines the rate and the epoch of the transition between the asymptotic region with $l_{dis} \propto (1+z)^{-\frac{1}{2}}$ and that with $l_{dis} \propto (1+z)^{-1}$.

We can now substitute into the above equations, (3.5) and (3.16), the transfer function for a model and compute the values of $l_0$, $z_t$ and $l_{LSS}$. We, thus, show in Figure 2 the results for the BBKS CDM, the BBKS WDM with $M_{part} = 3\,\text{keV}$, the Klypin et al. (1993) CHDM, the Gottlöber et al. (1994) BSI and the BBKS HDM power spectra.

Of specific interest is, of course, the scale for the present epoch, $z = 0$. We find for CDM (BBKS)

$$l_0 = 6.8 h^{-2} Mpc, \quad 1 + z_t = 1.1, \quad l_{LSS}(0) = l_{dis}(0) \approx 13.2 \text{Mpc} \quad (3.17a)$$

and for the different CDM transfer function of Bond & Efstathiou (1984)

$$l_0 = 7.15 h^{-2} Mpc, \quad 1 + z_t = 1.15, \quad l_{LSS}(0) = l_{dis}(0) \approx 14 \text{Mpc}. \quad (3.17b)$$

For the WDM spectrum BBKS with $M_{part} = 3\,\text{keV}$, we obtain

$$l_0 = 6.0 h^{-2} Mpc, \quad 1 + z_t = 1.6, \quad l_{LSS}(0) = l_{dis}(0) \approx 14.8 \text{Mpc}. \quad (3.17c)$$

For the CHDM spectrum (Klypin et al. 1993), we obtain

$$l_0 = 23.1 h^{-1} Mpc, \quad 1+z_t = 0.4, \quad l_{LSS}(0) = l_{dis}(0) \approx 11 \text{Mpc} \qquad (3.17d)$$

and for the BSI spectrum (Gottlöber et al., 1994)

$$l_0 = 39. h^{-1} Mpc, \quad 1+z_t = 0.15, \quad l_{LSS}(0) = l_{dis}(0) \approx 7.8 \text{Mpc}. \qquad (3.17e)$$

For the HDM spectrum (BBKS), $z_t > z_{cr}$ and we present $z_{cr}$ instead of $z_t$. Thus, in this case,

$$l_0 = 9.6 h^{-2} Mpc, \quad 1+z_{cr} = 0.45, \quad l_{dis}(0) \approx 13.2 \text{Mpc}, \qquad (3.17f)$$

The values for $1+z_t$ and $l_{LSS}(0)$ are for $h = 0.5$.

## 4 N-BODY SIMULATIONS

### 4.1 Measured and theoretical scales for simulated dark matter catalogues

With the specified and, thus, known characteristics of N-body simulations, it is natural to consider their use in illustrating the new theory developed here for an *approximate* quantitative description of the formation and evolution of large-scale structure. However, the catalogues of DM particles from such simulations are still of a 'discrete' distribution and there is, at present, no direct way of delineating Zel'dovich DM pancakes. So, as usual, some statistical method of analysis is required. But, most standard analyses have concerned themselves only with purely statistical measures of a stochastic point process or with various topological properties which then depend on some arbitrary smoothing parameter to transform the distribution of points into a smooth 'field' distribution. It was with the aim of tackling the problem of the *physical* measurement of large-scale structure that BDF proposed a new method of analysis, which we shall simply call the 'core-sampling' method. The method distinguishes between filament-like elements and sheet-like elements and in doing so also measures their characteristic scales. BDF applied it to actual observational catalogues, the deep pencil beam galaxy redshift surveys that were then available. We shall discuss the relevance of their results for this paper in Section 5.

Following BDF, DFM applied the 'core-sampling' analysis to a distribution of DM particles at the 'present epoch' from an N-body simulation with defined 'galaxies' (Eke et al. 1995). The analysis showed that $\sim 80$–$90\%$ of the particles were contained in the large-scale structure that had developed in the simulation. However, although filamentary structure dominates for the 'galaxy' distribution, as it also seems to do for actual galaxy catalogues (BDF), the situation for the DM particle distribution is entirely different. For DM, filaments are seen by the analysis for the poorest structure elements only and sheet-like structures for the greater part, i.e $\sim 70$–$80\%$, of the particles in the simulation. It seems then that the 'core-sampling' analysis clearly discerns the expected Zel'dovich pancake structure of the gravitational collapse of DM overdensities, with 'galaxies' in the simulations then tracing the 'ridges' of these DM pancakes. In particular, gravitational instability predicts then the presence of not just galaxy filaments, but also substantial halos of DM around the galaxy filaments. Clearly, with the poorest

Zel'dovich DM pancakes the 'core-sampling' analysis is only able to pick up their central 'ridges'.

Quantitatively, the filamentary network can be characterised by the surface density of filaments, $\sigma_f$, defined as the mean number of filaments crossing a unit area of arbitrary orientation. The sheet-like structure can be characterised by the mean separation, $D_s$, of the intersections of sheets with a random straight line. This is clearly also the mean linear intersection of a random cell with a random straight line. The linear density of sheets, $\sigma_s = 1/D_s$, evidently provides the same information.

Since the 'core-sampling' analysis 'finds' both filament- and sheet-like structures for DM LSS, we need, as a first step in our analysis, to define the mean separation of structure elements for a complex structure of both filaments and sheets. The simplest solution is to use the diameter of spheres, which when placed at random will on the average be intersected by two structure elements. As the mean number of 'planes' (i.e. sheets) and 'straight lines' (i.e. filaments) intersecting the sphere with radius $r$ is

$$< N_{el} > = 2\pi r^2 \sigma_f + 4\sigma_s r, \qquad (4.1)$$

we are led to define the mean separation, $D_{f-s}$, as

$$D_{f-s} \equiv 2\left(\sigma_s + \sqrt{\sigma_s^2 + \pi\sigma_f}\right)^{-1}. \qquad (4.2)$$

We shall not discuss here the intricacies of the analysis, as they are fully considered in Paper II, but we should note that, whereas in the observational data the background of galaxies has a negligible effect, for the DM particle catalogues the background may produce in our 'core-sampling' analysis a small, but significant, number of 'artificial' structure elements. That is, because of biasing through some threshold criterion for galaxy formation and through physical processes such as secondary ionization inhibiting formation, we might expect that *all* galaxies are incorporated into structure elements, which, of course, will not be the case for the DM. In particular, this makes the determination of the mean separation of DM structure elements quite sensitive to the richness criterion, as defined by BDF, with the *minimal* mean separation given by the lowest richness thresholds.

However, this does provide us with two types of *minimal* mean separation which can be estimated quite reliably. As the sheet-like elements are already richer structures, their mean separation, $D_s$, is then only weakly dependent on the values of the lowest thresholds. The other reliable quantity is the estimate, $D_{f-s}$, of equation (4.2) for the *full* sample of particles. Clearly, $D_{f-s}$ provides then a lower limit of the actual mean separation of DM LSS elements and $D_s$ an upper limit. The difference between these quantities will, of course, depend on how evolved the structure is, as earlier on less of the DM particles have condensed onto structure elements and the background is then greater.

We present in Figs 3(a)–(c) the comparison of the theoretical evaluations of the dissipative scale, equation (3.14), with the corresponding 'core-sampling' estimates of these minimal mean separations of LSS structure elements as measured for three simulations. The simulations were carried out for the following parameters:

1. $N_{cell}$, the number of cells along a side of the simulation cube, = 256, giving $256^3$ cells in total.

2. $N_p$, the number of particles, = $128^3$.

3. For the CDM-200 simulation in Fig. 3(a), the Bond & Efstathiou (1985) CDM spectrum was used with a computational box size of $L_{box}$ = $200\,h^{-1}$ Mpc.

4. The Broken Scale Invariance spectrum (BSI) of Gottlöber, Mücket & Starobinsky (1994) and Kates et al. (1995) was used for the other two simulations, with $L_{box} = 200\,h^{-1}$ Mpc for BSI-200 in Fig. 3(b) and $L_{box} = 25\,h^{-1}$ Mpc for BSI-25 in Fig. 3(c).

In these figures the dashed lines correspond to the theoretical function, $l_{LSS}(z)$, of equation (3.14) and the upright triangles and inverted triangles the 'core-sampling' estimates of $D_{f-s}$ and $D_s$, respectively. As we can see from Fig. 3(a), there is early structure formation for the CDM-200 model with theory predicting the redshift range for this phase to be $z \approx (5-10)$, with $0 < z < 2$ for the relatively late period of structure evolution, when structure disruption becomes significant. However, for the BSI-200 simulation theory predicts later structure formation with $z_{cr} \approx 2$. The dashed curves in all three models also clearly illustrate the rapidity of structure formation as implied by equation (3.5).

But, most encouraging for our work here is how the figures show that

1. Considering the approximate nature of our theory, there is excellent agreement between the redshift dependence given by equation (3.14) for $l_{LSS}(z)$ with that for the 'measured' scales over the dissipation phase of structure evolution in all three simulations.

2. It is also interesting that for the BSI-200 simulation both the 'measurements', $D_s$ and $D_{f-s}$, also begin to increase with redshift at $z = 2.5$ in accordance with the theoretical expectations. This region is, however, generally difficult for applying the 'core-sampling' analysis, as it corresponds to the phase of structure creation when the structure is not as pronounced as it is at $z = 0$.

3. The difference between the two 'measures', $D_s(z)$ and $D_{f-s}(z)$, is only about twice the error in the measures themselves, i.e. they differ by only $\sim 2\sigma$.

Thus, the results support our proposal for a *quantitative* theory of the formation and evolution of large-scale structure in the universe, at least for LSS structure. As we discuss in the next section, present simulations are as yet not advanced enough to be able to provide a comprehensive test of the theory for SLSS structure. Nevertheless, it is encouraging that there is such close agreement of the theoretical predictions with the 'measurements' of the characteristic scales of LSS in these N-body simulations over the wide range of epochs for which the 'core sampling' analysis can be applied with confidence. Of course, these comparisons are still somewhat limited and we shall be testing further our theory with a more representative sample of simulations in Paper II. Our purpose here is simply to illustrate the expected correspondence between the theoretical definitions of physical measures of large-scale structure and actual 'physical measurements' in N-body simulations.

### 4.2 On the limitations of simulations

Clearly, since the *actual* power spectrum of a simulation has, inevitably, cut-offs, $k_{min}$ and $k_{max}$, imposed by the finite box size and finite resolution, our theoretical discussion has serious implications for N-body simulations. Not only does this limitation with the simulations produce for the largest wavelengths large statistical variations between different simulations of the same physical model, but, as well as distorting both the matter and 'galaxy' distributions, it will also affect the evolution of large-scale structure. For the fundamental physical quantities that we consider in this paper, this is clear just from an examination of equations (3.1), (3.2), (3.5) and (3.14). For a quantitative assessment, we need to evaluate the integrals not with the 'true' power spectrum, but with an *effective power spectrum* that is limited between the $k_{min}$ and $k_{max}$ defined by the simulation's box size and resolution, respectively. We have, thus, in this Section numerically calculated the effect on the evolution of LSS by truncating the power spectra, that is, by simply integrating the 'true' power spectrum between $k_{min}$ and $k_{max}$, instead of between 0 and $\infty$.

Figs 4(a) and (b) show the expected effect of a finite resolution, i.e. of $k_{max}$, on the characteristic scale of LSS, $l_{LSS}$, as a function of $z$ for numerical simulations of the CDM and HDM models of BBKS. For these calculations, we chose a very large box size of $L_{box} = 6000\,h^{-1}$ Mpc, which to all intents and purposes amounts to an 'infinitely' large box. The resolution is then taken to be the cell size, in which case it is just $L_{box}/N_{cell}$, where $N_{cell}$ is, as usual, just the number of cells along a side of the simulation cube, i.e. such a simulation has $N_{cell}^3$ cells. Thus, these parameters give simulations, for which the *effective power spectra* will be limited to the range, $k_{min} \leq k \leq k_{max}$, with $k_{min} = 2\pi/L_{box}$ and $k_{max} = N_{cell}k_{min}$. To investigate the effect of different resolutions we have chosen $N_{cell} = 2^8$, $2^{10}$, $2^{12}$ & $2^{15}$, although in Fig. 4(a) only the results for $N_{cell} = 2^8$ & $2^{15}$ are shown, since for the HDM model there is little difference in the results for these values of $N_{cell}$. These then correspond to resolutions of 23.4, 5.9, 1.46 & 0.18 $h^{-1}$ Mpc, respectively. The spectra were all normalized to the first year COBE quadrupole amplitude of $17\,\mu K$ for the angular fluctuations of the cosmic microwave background (Smoot et al. 1992).

Fig. 4(c), on the other hand, shows the effect of using different box sizes for simulations of the (standard) CDM model of BBKS. For these results, a fixed $N_{cell}$ of 256 was used and simulations considered for the range of box sizes in regular use for present numerical simulations, that is, for $L_{box} = 25, 100, 300$ & $500\,h^{-1}$ Mpc, giving resolutions of 0.10, 0.39, 1.17 & 1.95 $h^{-1}$ Mpc, respectively. For comparison, the solid curve from Fig. 4(b) for a CDM simulation with $L_{box} = 6000\,h^{-1}$ Mpc and $N_{cell} = 2^{15}$ is also shown in Fig. 4(c), again as a solid curve.

Again, the rapid creation of structure at $(1+z) \sim 2(1+z_{cr})$, due to the exponential term in equation (3.5), is clearly depicted in all three figures. For almost all the models with $z_{cr} > 2$, the part of the curves depicting the merging or dissipation of LSS is indeed well fitted by equation (3.16) over the full range of epochs for these results; we remind the reader that (3.16) was proposed solely on the asymptotic behaviour with redshift of large and of small $l_{LSS}$. In particular, Fig. 4(c) shows explicitly that, except for the model with the smallest $L_{box}$ of $25\,h^{-1}$ Mpc, they do, indeed, all have the same scale-free asymptotic behaviour, *viz*,

$$l_{LSS} \propto (1+z)^{-1/2}, \qquad \text{for } z \text{ small},$$

corresponding to the fact that they all have the same long wavelength spectrum, which here is simply Harrison-Zel'dovich.

Clearly, the different shapes of the various spectra at short wavelengths lead to very different time dependences, Figs 4(a) and (b). For the HDM model, it is well known that structure forms much too late and, indeed, Fig. 4(a) shows that in this case the creation of new structure dominates up to an epoch with $1 + z \approx 0.25$. But, what may be more worrying for the comparison of the results of numerical simulations with observations is that the same situation also occurs for simulations of models with CDM spectra using a simulation box with $N_{cell} \lesssim 2^8$ in conjunction with $L_{box} = 6000\,h^{-1}$ Mpc, Fig. 4(b). Formally, the $k_{max}$ induced by such a poor resolution scale produces a power spectrum that is *effectively* HDM. Clearly, the same will also be true for similar simulations for CHDM spectra. Physically, what has happened is simply that there is then no information in the simulations about the evolution of structure below this scale, which would be the dominant process at early times. Instead, only the much larger scale structures are all that can be seen in such simulations.

In Table I, we present our results for the characteristic scale of LSS at $z = 0$, as well as the characteristic scale of SLSS and the critical epochs for the evolution of structure. Surprisingly, despite the evolutionary histories of LSS being so very different, the characteristic scales at $z = 0$ are very similar for most of the models. This, at least, is encouraging news.

It is for those current simulations using small box sizes that even the scale for LSS is incorrect, as we can see from the curve in Fig. 4(c) for the case of $L_{box} = 25\,h^{-1}$ Mpc and $N_{cell} = 256$. The (long-dashed) curve lies about a factor of 2 in $l_{LSS}$ below the (solid) comparison curve over the dissipation era common to the two, i.e. $z \lesssim 8$. Thus, for example, it would be unreasonable to compare with observations the bulk motion on these scales for simulations based on such small computational box sizes, as that from the simulations will also be too small; this bulk motion and the dissipation scale are approximately proportional to each other, cf. equation (3.13). Also essential to note is the extremely small value of $L_{SLSS}$ for such a box size, which will seriously affect the dynamical evolution of the structures on scales in the very important range, $L_{box} > l > L_{SLSS}$.

However, current simulations with $N_{cell} = 256$ and using too large a box size also have their problems, as they lead to an epoch of structure formation in the simulations that is much too late. For example, for the biggest box with $L_{box} = 500\,h^{-1}$ Mpc, the dotted curve in Fig 4(c) shows that structure forms at a low redshift of $z \approx 3$. Thus, as actual structure in the Universe has formed before $z = 3$, the distribution of galaxies and matter in the simulated structure elements and how the distribution evolves will differ significantly from the observed distribution. Indeed, the situation is rather like that for the HDM power spectrum, although clearly not as bad.

Thus, although we can obtain good agreement with perhaps the most popular measure of large-scale structure, the correlation function, there are some important aspects of the simulations that can be seriously different from observation, even if they are based on a correct or approximately correct power spectrum. If we compare the curves in Fig. 4(c) with the solid comparison curve, it would seem that the best compromise box for present $N_{cell} = 256$ simulations of the formation and evolution of LSS in a CDM universe is one with $L_{box} \approx 100\,h^{-1}$ Mpc. Unfortunately, although it is then possible to simulate LSS, the results for SLSS may not be valid, as, clearly, such a box can only contain very few elements of SLSS. In which case, any conclusions one may wish to draw from such simulations about the formation and distribution of clusters of galaxies, say, must be viewed with some circumspection, as the large correlation length of clusters clearly show these to be elements of SLSS.

It seems then that there are severe limitations with present N-body simulations for a realistic comparison with observations, but used *with care* they can still provide a limited means of constraining theories for the formation and evolution of structure in the Universe.

## 5 SUMMARY, DISCUSSION AND CONCLUSIONS

### 5.1 Summary of results

Our more important results are:

1. The exponential term in equation (3.5) shows that the creation or formation of structure occurs with great rapidity, and much of the structure is in place within a short time interval at the redshift $z_{min}$, with $1 + z_{min} \approx 2.2(1 + z_{cr})$. It is, thus, particularly appealing to consider this process as a *phase transition* from a structureless universe to a structured one.

2. After this 'phase transition' is a particularly quiet, long-lasting, phase of structure *evolution,* during which two processes are dominant:
    a) The accretion of matter by the structure elements created during the phase transition, resulting in the growth of the fraction of matter in structure and in the formation of a 'broken structure' in the universe (Section 3.3; see, also, BDF).
    b) The merging of structure elements and the consequent growth of the characteristic scale of LSS. This process of dissipation is described by equations (3.14) and (3.16).

3. At the end of Section 3.2 we discussed the effect on $z_{cr}$ of the gravitational potential, which spatially modulates it on very large scales to produce a 'random cellular framework' to large-scale structure in the universe. The characteristic spatial scale of the modulation, i.e. the average 'cell' size, is defined by equations (3.1), (3.2) and (3.3). For any reasonable model this scale is large, $\gtrsim 50\,h^{-1}$ Mpc, and its comoving value is then constant with redshift.

4. The excellent agreement between the values resulting from the purely theoretical definitions of the characteristic scale for LSS and the 'measurements' of the 'mean separation' of LSS elements given by the 'core sampling' analyses of DM catalogues from N-body simulations over a wide range of redshifts gives strong support for the new theory that we propose here for a *quantitative* description of the formation and evolution of large-scale structure in the universe.

5. An immediate consequence of the theory is that N-body simulations have then severe limitations for the description of large-scale structure. For the study of the formation and evolution of LSS in a CDM universe, the best compromise for the $N_{cell} = 256$ used in present simulations is for a box size of $L_{box} \approx 100\,h^{-1}$ Mpc. However, the results for SLSS may then not be valid.

### 5.2 Theory and observations

Clearly, we can use our theory to address the question of large-scale structure in the DM distribution and, thus, it is immediately applicable to N-body simulations, as we have seen in Section 4. However, in Section 2 we presented a brief overview of the qualitative theoretical picture of the formation and evolution of structure in the universe and saw that galaxies are a 'biased' distribution relative to the dark matter. It is also clear then that, at least for LSS, the relationship between the two distributions will most likely need *realistic* physical simulations for its elucidation.

A first step is made using the 'thresholding' criterion of BBKS and, indeed, in a 'core-sampling' analysis of such a simulation with defined 'galaxies' (Eke et al. 1995) DFM did find that the 'galaxies' do not faithfully trace the DM distribution, but traced only the 'ridges' of Zel'dovich DM pancakes. However, secondary ionization will generate an even stronger bias. The Gunn-Peterson effect in QSOs at $z \approx 4.5$ (Schneider et al. 1991), suggest that secondary ionization may have occurred at $z \simeq 5$. This would explain, in particular, the extraordinary voids seen in galaxy redshift catalogues, with deeper catalogues serving mainly to emphasize structure already seen in shallower catalogues in the volume common to both. It is, thus, expected that some of the weak 'galaxy' filaments seen in present simulated 'galaxy' catalogues would not have any galaxies if secondary ionization had also been included. Indeed, this may explain the observations by Morris et al. (1993) of a relatively small group of Ly-$\alpha$ clouds, with one cloud some $6.6\,h^{-1}$ Mpc away from the nearest observed galaxy, as belonging not to the halos of galaxies, but rather to a Zel'dovich DM pancake within a void in the galaxy distribution. At present, one is only just beginning to contemplate how secondary ionization may be mimicked in a simulation. But, the crucial point to understand is that it is also abundantly clear that this will furthermore result in a bias that is different in over-dense regions from its value in under-dense regions. That is, secondary ionization will produce a bias that depends on environment and, thus, on position.

Therefore, until there are more *realistic* simulations, we can only apply our theory for the formation and evolution of LSS to observational data in a more qualitative way. In the CDM simulation analysed by DFM, we found $D_f^{'gal'} \approx 1.4\, D_{f-s}^{DM}$. But, with secondary ionization, we would expect that the difference between $D_f^{gal}$ and $D_{f-s}^{DM}$ could be appreciably greater. In fact, the comparison of the theoretically predicted values for $l_{LSS}$ at the present epoch, equations (3.17a–f), with the observational estimate of BDF of $\approx 14\,h^{-1}$ Mpc for their observational data, at least, suggests that the characteristic scale of the galaxy LSS is about a factor of two greater than that of DM LSS.

The form of the evolution of DM LSS, $l_{dis}(z)$, is clearly of great physical interest and we would hope that this is also the approximate form for the evolution of LSS in the distribution of galaxies. However, equations (3.17a–f) and Fig. 2 show that its value, as well as its dependence on redshift, during this quiet evolutionary phase are similar for the different models that we have considered here. This would suggest that the measurement of the characteristic scale of LSS for galaxies, although of intrinsic physical interest, may well not be an appropriate measure for discriminating between various models. Perhaps, this may not be much of a surprise, because if galaxy filaments are the dominant population, as observations would seem to show, the characteristic scale of LSS is strongly related to the coherence length of the galaxy correlation function (BDF) and it is well-known that the two-point correlation function is a rather insensitive measure.

Most interestingly, on the other hand, is our result that $z_{cr}$ is clearly sensitive to the models. But as this is intimately connected with the mass of the DM particles expected to dominate the evolution of the universe, this also may not be surprising. Thus, the fact that we can observe radio galaxies and QSOs to high redshifts places a strong constraint on possible theories for the formation of structure. In particular, the rapidity of the creation of structure shows that we must then have $z_{min} > 4$ or, equivalently, $z_{cr} > 1.5$. As is now well-known, it is not surprising then that the HDM model proved so unsatisfactory, and is now a model mainly of heuristic interest. Fig. 2 would also seem to rule out the CHDM model, as it seems unable to provide any structure formation before $z \approx 3$. However, this is not a fundamental defect of the model, but is rather a consequence of the analytical relationship which Klypin et al. (1993) had fitted to the power spectrum. Indeed, for this model the transfer function $T(k)$ has a zero at a finite $k$ value. Such a zero in $T(k)$ cannot be explained by any physical effect and, thus, a better fit would be required if we wish to estimate the epoch of structure formation for this model.

From the general picture presented in Section 2, it is clear that the characteristic scale for SLSS may well be a robust measure. Indeed, with the possibility of secondary ionization 'enhancing' or 'high contrasting' the SLSS caused by the modulation of the velocity field by the gravitational potential, it would seem that one would actually be more successful in this case of obtaining the SLSS scale for large-scale structure by analysing the galaxy distribution instead of the DM distribution, even if the latter were possible.

Thus, it is interesting that, with the galaxy pencil beam redshift surveys then available, BDF found a characteristic scale for SLSS of $L_{SLSS} \sim 50\,h^{-1}$ Mpc for the better sampled deep pencil beam surveys. A scale of this order does seem to be further supported by a visual inspection of the three-dimensional galaxy distributions reported from much larger galaxy redshift catalogues (Shectman et al. 1994, Ratcliffe et al. 1995). Even with such an inaccurate estimate, the theoretical predictions for the models we have considered here are different enough to make it of more than passing interest to compare them with observation. It would seem that the CDM model is then in best accord with the present observational results. Of course, it is premature at this stage to be drawing such inferences and the analysis needs to be repeated using these 'fairer samples' of the Universe that are now being acquired.

Finally, even without these problems for the *realistic* simulation of galaxy catalogues, it is clear from Section 4 that the direct comparison of the results of N-body simulations with observations should be treated with caution. The necessarily truncated character of the initial power spectrum that a finite simulation requires will lead to the distortion of the characteristic scales of large-scale structure. Thus, there can be severe limitations as to how well simulated catalogues can reproduce all the actual parameters of the distribution of galaxies in the Universe. This simple fact demonstrates once again the importance of obtaining even an approximate purely theoretical description of the evolution of structure.

### 5.3 Conclusions

In this paper we have defined a quantitative theory for an approximate description of the formation and evolution of structure in the universe. This is the first time that there has been such a *quantitative* theory. Just as the Zel'dovich approximation established the *general* behaviour of the non-linear evolution of density inhomogeneities in the post-recombination universe and predicted the phenomenon of large-scale structure in the universe, the theoretical discussion of Section 3, which can also be thought of as an extension of Zel'dovich's theory, establishes the general properties of the first two phases of the formation and evolution of large-scale structure. Our more qualitative results are similar to those obtained by the adhesion approximation (KPSM), but we re-emphasize again that our aim here has been to provide a more detailed *quantitative* theory. This has been achieved through our 'statistical mechanical' approach to the problem as opposed to the 'fluid dynamical' approach of the Zel'dovich approximation and of the adhesion approach (Section 3.4). Thus, most importantly, our theoretical definitions for various physical measures of structure provide this quantitative description. In particular, we define the physical scales of both LSS and SLSS. The successful comparison of these theoretical values with 'measured' values of the 'mean separations' of LSS as found through the 'core sampling' analysis of N-body simulations support our claim that our theoretical definitions do correspond to physical measures of the distribution of matter in the universe.

We, of course, need to test and, perhaps, refine further the theory and our method of analysis of simulated DM and real or simulated galaxy catalogues much more extensively using N-body simulations (Paper II) and the more representative observational catalogues that have just been acquired, such as the Durham/UKST redshift survey (Ratcliffe et al. 1995) and the Las Campanas redshift survey (Shectman et al. 1994). However, the full confrontation of theory with observations can only happen when in the distant future we have reliable simulations of *real* galaxies, that is, dynamical simulations that give a good account of the physical criteria for the formation of galaxies including the position dependent biasing that would be caused by secondary ionization. But, clearly, the results of this paper demonstrate, at least, the potential power of our approach for the understanding of how large-scale structure in the Universe may have formed and evolved.

## Acknowledgements

AGD wishes to acknowledge support from the Center of Cosmo-Particle Physics, Moscow and support by the Danmarks Grundforskningfond through its establishment of the Theoretical Astrophysics Centrum. AGD, RF and SG would also like to acknowledge INTAS for support for travel in connection with this work. We would like to thank Shaun Cole for supplying the estimates of the bulk velocities in their simulations and Marek Demiański and Sergei Shandarin for useful discussions.

**Figure Captions**

**Figure 1.** Comparison of theoretical calculations, as given by the curve, with estimates of the bulk velocity of dark matter from two different realizations of an N-body simulation with a CDM power spectrum, depicted as filled circles and triangles.

**Figure 2.** Theoretical curves for the dependence of the scale of LSS with redshift for CDM, WDM, BSI and CHDM power spectra (see text).

**Figure 3.** The characteristic scale of LSS, $l_{LSS}(z)$, for (a) the CDM-200, (b) the BSI-200 and (c) the BSI-25 N-body simulations according to equation (3.15). Filled upright and inverted triangles correspond to the two different estimates of the 'mean separation' of LSS structure elements, $D_{f-s}$ and $D_s$, respectively, using the 'core sampling' method of analysis (see text).

**Figure 4.** The characteristic scale of LSS, $l_{LSS}(z)$, for the (a) HDM and (b) CDM N-body simulations according to equation (3.15) for $L_{box} = 6000\,h^{-1}$ Mpc and for different shortwave cutoffs; for the HDM simulations $N_{cell} = 2^8$ & $2^{15}$ and for the CDM simulations $N_{cell} = 2^8, 2^{10}, 2^{12}$ & $2^{15}$. (c) Similarly, $l_{LSS}(z)$ for CDM simulations with the same value for $N_{cell} = 256$, but now with $L_{box} = 500, 300, 100$ & $25\,h^{-1}$ Mpc. For comparison, the full curve for $L_{box} = 6000\,h^{-1}$ Mpc and $N_{cell} = 2^{15}$ is also presented and is the same as the full curve in (b).

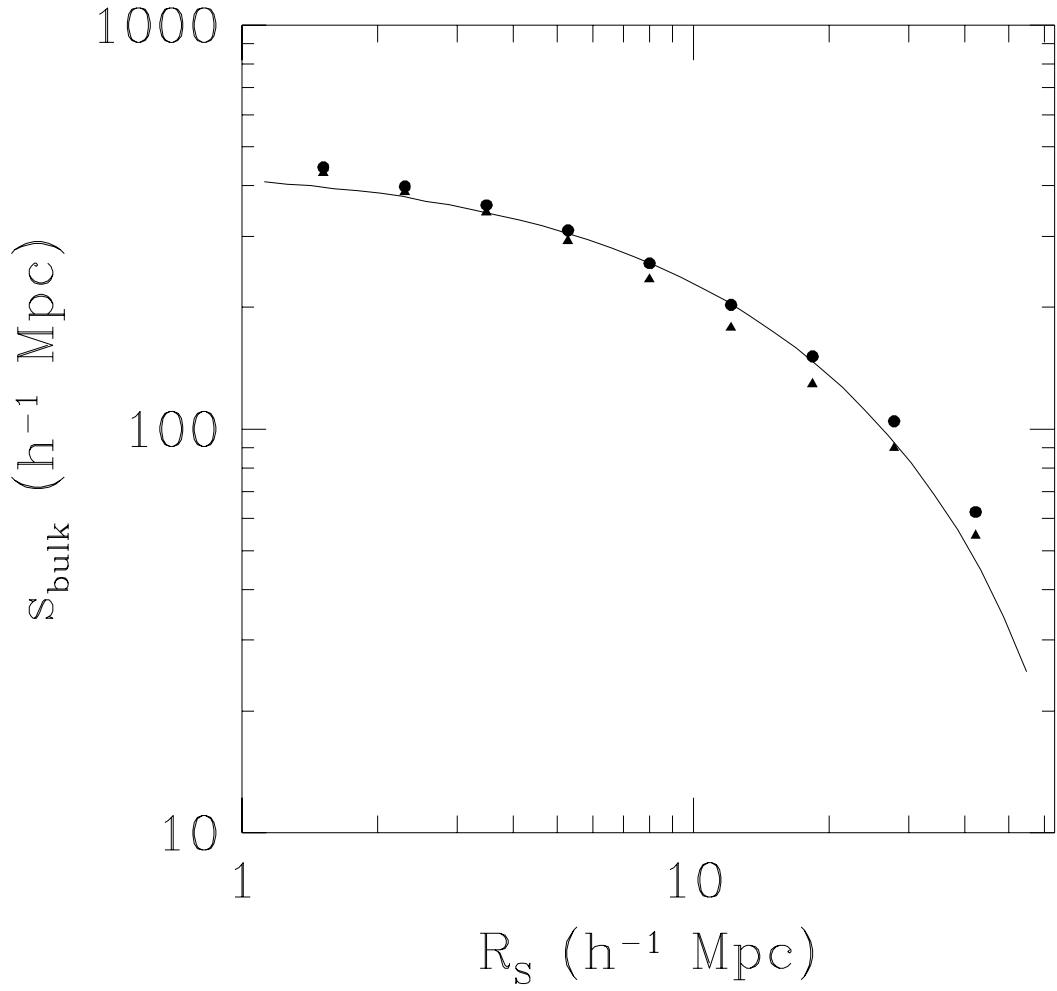

Fig. 1

Table I. Theoretical values for the characteristic scales of SLSS and LSS at $z=0$ and for $z_{cr}$, $z_t$ and $z_{min}$ for the CDM and HDM models

| MODEL | $L_{box}$ $h^{-1}$Mpc | $N_{cell}$ | $L_{SLSS}$ $h^{-1}$Mpc | $l_{LSS}$ $h^{-1}$Mpc | $1+z_{cr}$ | $1+z_t$ | $1+z_{min}$ |
|---|---|---|---|---|---|---|---|
| HDM | 6000 | 32000 | 110. | 85. | 0.5 | 0.6 | 0.2 |
| HDM | 6000 | 256 | 112. | 120. | 0.4 | 0.6 | 0.2 |
| CDM | 6000 | 32000 | 60. | 6.4 | 8.3 | 1.1 | 17.4 |
| CDM | 6000 | 4096 | 61. | 6.4 | 4.5 | 1.1 | 6.4 |
| CDM | 6000 | 1024 | 64. | 6.5 | 2.5 | 1.1 | 2.1 |
| CDM | 6000 | 256 | 75. | 13.2 | 1.0 | 0.9 | 0.5 |
| CDM | 500 | 256 | 54. | 6.4 | 4.0 | 1.0 | 5.3 |
| CDM | 300 | 256 | 47. | 6.0 | 4.8 | 1.0 | 7.2 |
| CDM | 100 | 256 | 25. | 4.5 | 6.8 | 0.7 | 11.8 |
| CDM | 25 | 256 | 8. | 2.0 | 9.6 | 0.2 | 17. |

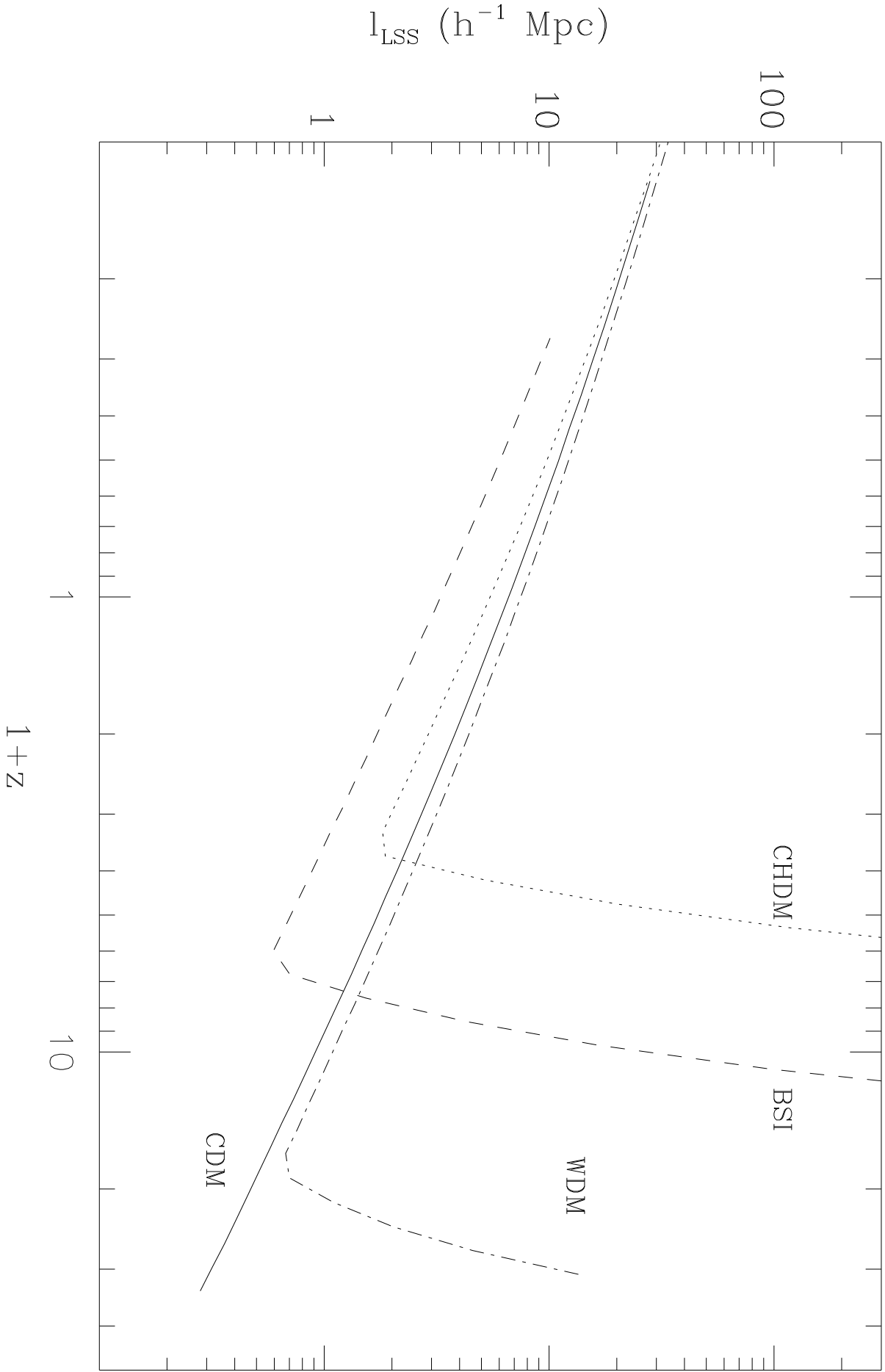

Fig. 2

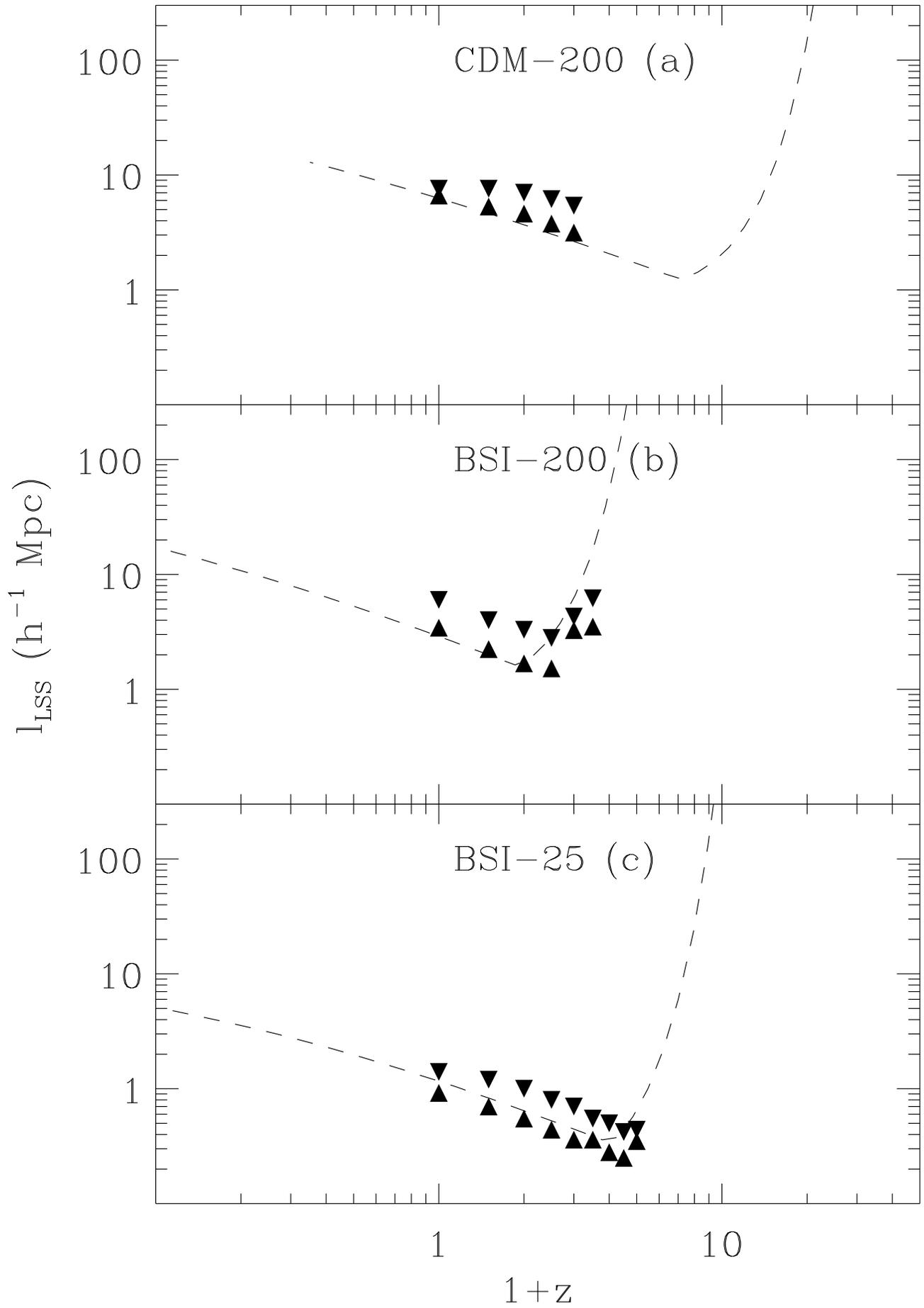

Fig. 3

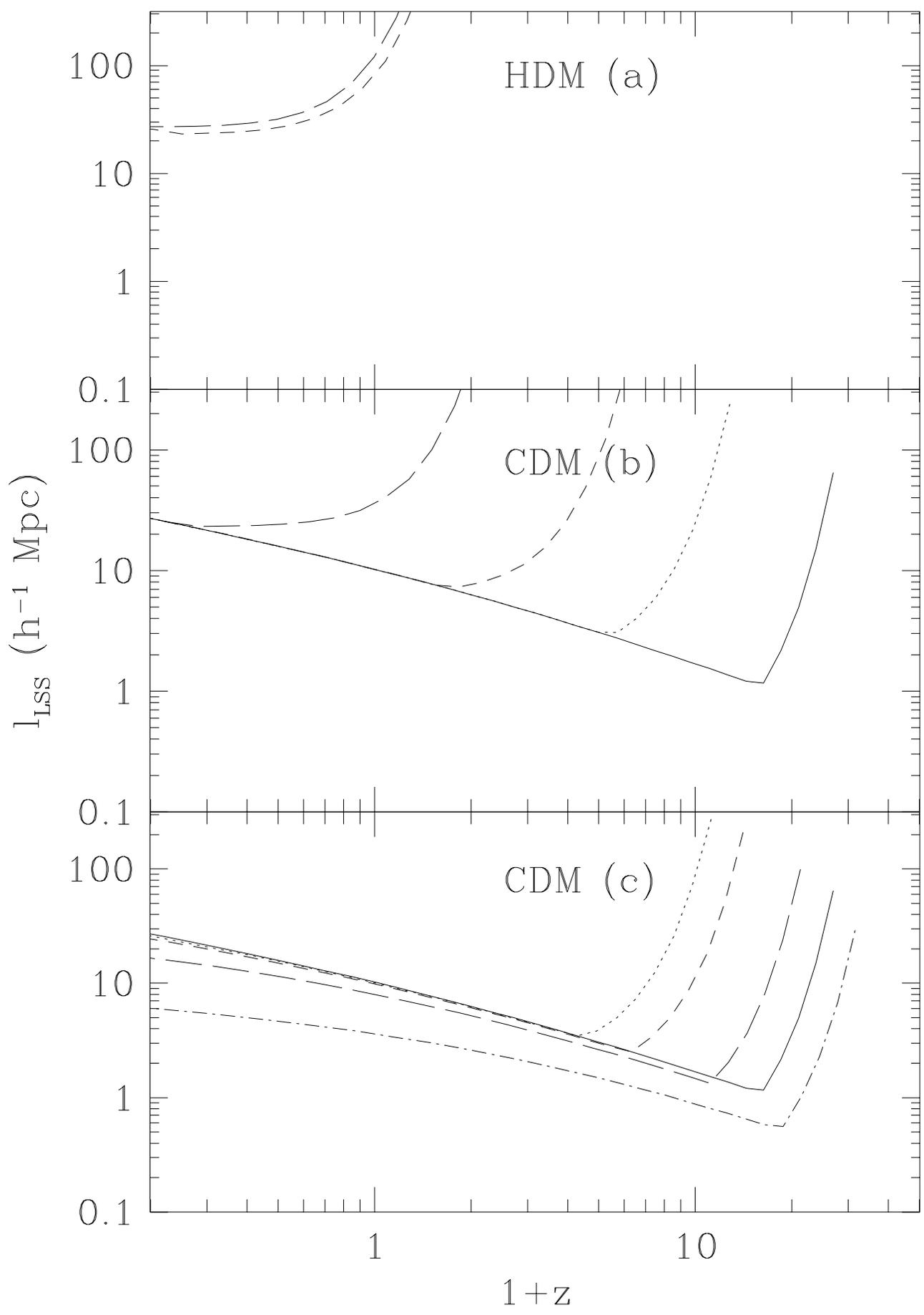

Fig. 4